\documentclass[authoryear,final,5p,times,twocolumn]{elsarticle}
\usepackage{bm}
\usepackage{txfonts}
\usepackage[T1]{fontenc}
\usepackage[colorlinks=true,citecolor=blue,linkcolor=blue,urlcolor=blue,breaklinks]{hyperref}   
\usepackage{cleveref}
\usepackage{aas_macros}                    
\usepackage{amssymb}                    
\usepackage{enumerate}                  
\usepackage{float}                      
\usepackage{natbib}                     
\usepackage{textcomp}                   %
\usepackage{xcolor}					   
\usepackage{footnote}                   
\usepackage{textcomp}                   
\usepackage{tabularx}                   
\usepackage{threeparttable} 			   
\usepackage{multirow}                   
\usepackage{booktabs}                   
\usepackage{framed}                     
\usepackage{bm}                         
\usepackage{graphicx}                   
\usepackage{enumitem}                   
\usepackage{subfigure}                  
\usepackage{rotating}
\usepackage{pdflscape}

\makeatletter
\renewcommand\appendix{\par
  \setcounter{section}{0}%
  \setcounter{subsection}{0}%
  \setcounter{equation}{0}%
  \setcounter{table}{0}
  \setcounter{figure}{0}
  \gdef\theequation{\@Alph\c@section.\arabic{equation}}%
  \gdef\thefigure{\@Alph\c@section.\arabic{figure}}%
  \gdef\thetable{\@Alph\c@section.\arabic{table}}%
  \gdef\thesection{\appendixname\@Alph\c@section}%
  \@addtoreset{equation}{section}%
  \@addtoreset{table}{section}
  \@addtoreset{figure}{section}
}
\makeatother

\def\klm{}
\def\klmr{}

\crefname{appsec}{Appendix}{Appendices}
\journal{Astronomy \& Computing}
\begin{document}
\begin{frontmatter}

\title{The Apertif Monitor for Bursts Encountered in Real-time (AMBER) auto-tuning optimization with genetic algorithms}

\author[uva,astron]{K.~Mikhailov\corref{cor1}}
\ead{K.Mikhailov@uva.nl}
\author[nlesc]{A.~Sclocco}
\ead{a.sclocco@esciencecenter.nl}
\cortext[cor1]{Corresponding author}
\address[uva]{Anton Pannekoek Institute for Astronomy, University of Amsterdam, Science Park 904, 1098 XH Amsterdam, The Netherlands}
\address[astron]{ASTRON, the Netherlands Institute for Radio Astronomy, Postbus 2, 7990 AA, Dwingeloo, The Netherlands}
\address[nlesc]{NLeSC, Netherlands eScience Center, Science Park 140, 1098 XG Amsterdam, The Netherlands}

\begin{abstract}
Real-time searches for faint radio pulses from unknown radio transients are computationally challenging. Detections become further complicated \klm{due to continuously increasing technical capabilities} of transient surveys: telescope sensitivity, searched area of the sky, number of antennas or dishes, temporal and frequency resolution. \klm{The new Apertif transient  survey} on the Westerbork telescope happens in real-time on GPUs by means of the single-pulse search pipeline AMBER~\citep{Sclocco-2017}. AMBER initially carries out auto-tuning: it finds the most optimal configuration of user-controlled parameters per each of \klm{four} pipeline kernels so that each kernel performs its task as fast as possible. The pipeline uses a brute-force (BF) exhaustive search which in total takes 5 -- 24 hours to run depending on the processing cluster architecture. We apply more heuristic, biologically driven genetic algorithms (GAs) to limit the exploration of the total parameter space, \klm{tune all four kernels together} and reduce the tuning time to few hours. Our results show that \klmr{after only few hours of tuning, GAs always find similar or even better configurations for all kernels together than the combination of single kernel configurations tuned by the BF approach}. At the same time, by means of their genetic operators, GAs converge into better solutions than those obtained by pure random searches. The \klm{explored} multi-dimensional parameter space is very complex and has multiple local optima as the evolution of randomly generated configurations does not always guarantee global solution.
\end{abstract}

\end{frontmatter}

\section{Introduction}
Various radio transient surveys constantly search for new pulsars, rotating radio transients~\citep[RRATs,][]{McLaughlin-2006}, and fast radio bursts~\citep[FRBs,][]{Lorimer-2007, Petroff-2016}, especially at \klm{less explored} extragalactic distances in dense environments. More such discoveries can help us better classify transients and study intergalactic medium (IGM). \klm{Even though distant radio transients are hard to localize, better localization} can more comprehensively explore Galactic and extragalactic source populations in terms of stellar evolution and star formation that should depend on the type of host galaxy.

\klm{New} discoveries of \klm{single} bursts with Parkes, UTMOST, and ASKAP~\citep{Caleb-2017, Bannister-2017, Bhandari-2018} and one repeating source of bursts with Arecibo~\citep{Spitler-2016, Chatterjee-2017} \klm{reveal new properties} of radio bursts. \klm{Searches for much fainter and more distant bursts require more fine-grained searches and lead to new processing challenges~\citep{Magro-2011, Barsdell-2012, Sclocco-2016}}. Just like standard pulsar searches, transient lookups are performed in the two-dimensional, time-frequency space for every unit of dispersion measure (DM, third dimension). Modern searches (see Table~\ref{tab:surveys}) are performed in real-time \klm{to trigger multi-frequency follow-up. They also require very high time sampling and frequency resolution to better determine the burst structure}. Growing data rates and computational costs require larger supercomputers and search pipelines based on graphics processing units (GPUs) rather than central processing units (CPUs)\footnote{Other options, such as FPGAs and ASICs, are also available. However, FPGAs are very hard to program, and floating point performance is not comparable with GPUs, whereas ASIC are expensive to design and produce.}.

\begin{savenotes}
\begin{table*}[t!]
\centering
\caption{Modern radio pulsar and transient surveys and their main characteristics. FoV is a survey field of view in square degrees, N$_\mathrm{beam}$ is a number of facilitated beams, t$_\mathrm{samp}$ is a sampling resolution in micro-seconds, $n_\mathrm{pol}$ is a number of polarizations, $\nu_\mathrm{cntr}$ and $\Delta\nu$ are central frequency and available bandwidth, both in megahertz, $T_\mathrm{sys} / G$ and $S_\mathrm{min}^\mathrm{3\,ms}$  are system noise and minimum detectable flux density for a $10\,\sigma$ single-pulse threshold and $3\,\mbox{ms}$ pulse width, both in Janskys.}
\begin{threeparttable}
\scalebox{0.8}{
\begin{tabular}{c c c c c c c c}
\hline\hline
\textbf{Parameter} & \textbf{CHIME}~\tnote{a} & \textbf{UTMOST}~\tnote{b} & \textbf{SUPERB}~\tnote{c} & \textbf{ASKAP}~\tnote{d} & \textbf{ALERT}~\tnote{e} & \textbf{SKA-Low}~\tnote{f} & \textbf{SKA-Mid}~\tnote{f} \\
\hline
\klm{Status} & \klm{commissioning} & \klm{ongoing} & \klm{ongoing} & \klm{ongoing} & \klm{commissioning} & \klm{future} & \klm{future} \\
FoV (deg$^2$) & 220 & 9 & 0.6 & 30 & 8.7 & 27 & 0.49 \\
N$_\mathrm{beam}$ & 1024 & 352 & 13 & 288 & 2600 & 500 & 1500 \\
t$_\mathrm{samp}$ ($\mu$s) & 2.5 & 655.36 & 64	& $-$ & 40.92 & 50 & 50 \\
n$_\mathrm{pol}$ & 2 &	1 & 4 & 2 & 2 & 4 & 4 \\
$\nu_\mathrm{cntr}$ (MHz) & 600 & 835.5 & 1382 & 1400 & 1400 & 250 & 800 \\
$\Delta\nu$ (MHz) & 400 & 31.25 & 400 & 336	 & 300 & 100 & 300 \\
N$_\mathrm{chan}$ & 1024 & 320 & 1024 & 336 & 1536 & 8192 & 4096 \\
$T_\mathrm{sys} / G$ (Jy) & 45 & 28.5 & 60 & 1800 & 70 & $2\times10^{-5}$ & $7\times10^{-6}$ \\
$S_\mathrm{min}^\mathrm{3\,ms}$ (SP, Jy) & 0.25 & 0.9 & 0.3& 13.4 & 1.6 & $1.8\times10^{-7}$ & $3.7\times10^{-8}$ \\
\end{tabular}
}
\begin{tablenotes}
\vspace{0.5cm}
\small
\item[a] \klm{Based on the CHIME system overview~\citep{Amiri-2018}}
\item[b] Based on the UTMOST system overview~\citep{Bailes-2017, Caleb-2017}
\item[c] Based on the SUPERB survey overview~\citep{Keane-2018, Bhandari-2018}
\item[d] Based on ASKAP survey description~\citep{Bannister-2017}
\item[e] Based on Apertif Incoherent Search setup~\citep{Maan-2017}
\item[f] Based on the updated SKA review~\citep{Dewdney-2013, Braun-2015, Levin-2017}
\end{tablenotes}
\end{threeparttable}
\label{tab:surveys}
\end{table*}
\end{savenotes}
\klm{Transient surveys are also technically limited in their ability to detect new bursts}. The number of antennas or dishes in the survey relates to the corresponding amount and size of beams they can produce. This determines how large the observing area of the sky would be. \klm{The system equivalent flux density $S_\mathrm{sys} = T_\mathrm{sys} / G$, where $T_\mathrm{sys}$ is a total system temperature and $G$ is a system gain. Together with frequency bandwidth $\Delta\nu$, number of polarizations $n_\mathrm{pol}$, single-pulse threshold and single-pulse width, this determines down to what extent of radio transient brightness we can possibly search~\citep[radiometer equation,][]{LorimerKramer-2004}}. Finally, \klm{the temporal and frequency resolution of the instrument set the limits to which the intrinsic structure of the pulse can be studied}. Within all such limitations, the data should be optimally distributed on CPUs and GPUs for the signal processing: this includes de-dispersion (appropriate shift and integration of frequency channels that removes frequency \klmr{dispersion}), signal smoothing, and signal-to-noise evaluation. One way to find configurations that allow for fast data distribution and processing is to perform auto-tuning. In this case every configuration gets tested in terms of the best possible performance (fastest processing time in case of radio transient surveys). In the end, the most optimal configuration that allows the fastest search gets chosen. Auto-tuning is widely applied in computer science~\citep{Williams-2008}, but has also seen applications in other domains such as computational finance~\citep{Gray-2013} or astronomy~\citep{Sclocco-2012}. Auto-tuning for radio transient surveys also shows promising results in terms of performance portability~\citep{Sclocco-2015}.

\klm{The new real-time Apertif survey on the Westerbork (WSRT) telescope (ALERT, the Apertif Lofar Exploration of the Radio Transient Sky\footnote{\url{http://alert.eu}}) is now equiped with a new 160$\times$GPU cluster that achieves 1.3\,Pflops of peak performance and a data rate of 4\,Tbit/s, and has 2\,PBytes of available storage space~\citep{{Maan-2017}}}. Such computational capacity enables deep searches up to $42\,\mu\mbox{s}$ time and $0.195\,\mbox{MHz}$ frequency resolution, respectively. Apertif front-ends on 12 WSRT dishes produce more than 400 tied array beams that in total cover 8.7\,deg$^2$ of the sky, searched between 1100-1750\,MHz with a tunable bandwidth of 300\,MHz. Commissioning data from a targeted search toward FRB121102 already suggested a detection~\citep{Oostrum-2017}.

All hardware and software constraints require an optimized distribution of processing resources on the cluster to allow for the fastest real-time search: GPU threads and items, local memory re-use, loop transformations. Auto-tuning allows for an automated search of these parameters. Although such tuning is performed only once for a running survey, it should be invoked again in case the survey undergoes hardware changes (e.g. front-end or back-end upgrades) or the search pipeline itself gets extended or improved (e.g. \klm{by adding new processing steps}). Besides, it should be easily portable to any other survey pipelines. 

Section~\ref{sec:amber} introduces the current search pipeline for ALERT and its current auto-tuning. We introduce a more heuristic approach for auto-tuning with genetic algorithms in Section~\ref{sec:ga}. Section~\ref{sec:results} shows achieved performance based on different algorithm input parameters as well as comparison with the pure random search. We discuss auto-tuning parameter space in terms of complexity and degeneracy in Section~\ref{sec:discussion} and draw our conclusions in Section~\ref{sec:final}. 

\section{AMBER auto-tuning}\label{sec:amber}
The real-time search for new single bursts on WSRT is performed via the single-pulse search pipeline AMBER~\citep[The Apertif Monitor for Bursts Encountered in Real-time\footnote{\url{https://github.com/AA-ALERT/AMBER/}},][]{Sclocco-2017}. The pipeline can be divided into four main operations or kernels: a two-step de-dispersion\footnote{For a single DM, the frequency channels are first united into subbands such that the radio pulse signal first gets de-dispersed along subbands (step one, subband de-dispersion), and then within each subband (step two, intra-subband de-dispersion).}, de-dispersed time series downsampling (smoothing) and subsequent signal-to-noise (S/N) computation. Before the search, each kernel of the pipeline gets tuned to find its most optimal processing configuration\footnote{\url{https://github.com/AA-ALERT/AMBER_setup/tree/ARTS_tender}}. 

\begin{savenotes}
\begin{table*}[t!]
\centering
\caption{Fixed survey parameters and their values for ARTS0 GPUs.}
\begin{threeparttable}
\scalebox{0.8}{
\begin{tabular}{c c c}
\hline\hline
\textbf{Parameter} & \textbf{Description} & \textbf{Value for ARTS0} \\
\hline
\texttt{DEVICE$\_$PADDING} & Size of the cache line of OpenCL device (bytes) & 128\tnote{*}  \\
\texttt{DEVICE$\_$THREADS} & Number of simultaneously running OpenCL work-items & 32\tnote{*}  \\
\texttt{MIN$\_$THREADS} & Minimum number of OpenCL work-items & 8 \\
\texttt{MAX$\_$THREADS} & Maximum number of OpenCL work-items & 1024\tnote{*} \\
\texttt{MAX$\_$ITEMS} & Maximum number of variables which the automated code is allowed to use & 255\tnote{*} \\
\texttt{LOCAL} & Use of OpenCL local memory & "-local" \\
\texttt{MAX$\_$ITEMS$\_$DIM0} & Maximum number of OpenCL work-items in time dimension & 64 \\
\texttt{MAX$\_$ITEMS$\_$DIM1} & Maximum number of OpenCL work-items in DM dimension & 32 \\
\texttt{MAX$\_$DIM0} & Maximum number of OpenCL work-groups in time dimension & 1024 \\
\texttt{MAX$\_$DIM1} & Maximum number of OpenCL work-groups in DM dimension & 128 \\
\texttt{MAX$\_$UNROLL} & Maximum loop unrolling & 32 \\
\texttt{INPUT$\_$BITS} & Processing data rate (bits per sample) & 8 \\
\texttt{SUBBANDS} & Number of frequency subbands & 32 \\
\texttt{SUBBANDING$\_$DMS} & Number of DM subbands & 2048 \\
\texttt{SUBBANDING$\_$DM$\_$FIRST} & Initial DM of the first subband (pc/cc) & 0.0 \\
\texttt{SUBBANDING$\_$DM$\_$STEP} & Subband DM step (pc/cc) & 2.4 \\
\texttt{DMS} & Number of DMs within each subband & 24 \\
\texttt{DM$\_$FIRST} & Initial DM within the first subband (pc/cc) & 0.0 \\
\texttt{DM$\_$STEP} & DM step (\texttt{SUBBANDING$\_$DM$\_$STEP} / \texttt{DMs}) (pc/cc) & 0.1 \\
\texttt{BEAMS} & Number of compound beams & 1 \\
\texttt{SYNTHESIZED$\_$BEAMS} & Number of synthesized beams & 1 \\
\texttt{MIN$\_$FREQ} & Minimum observing frequency (MHz) & 1290 \\
\texttt{CHANNELS} & Number of frequency channels & 1536 \\
\texttt{CHANNEL$\_$BANDWIDTH} & Frequential resolution (MHz) & 0.1953125 \\
\texttt{SAMPLES} & Number of samples & 25600 \\
\texttt{BATCHES} & Number of samples per chunk of data  & 10 \\
\texttt{SAMPLING$\_$TIME} & Time resolution ($\mu$s) & 0.00004096 \\
\texttt{DOWNSAMPLING} & Downsampling factors & 10 $\dots$ 3200 \\
\texttt{NRSAMPLES} & Downsampled number of samples & \texttt{SAMPLES} / \texttt{DOWNSAMPLING} \\
\end{tabular}
}
\begin{tablenotes}
\vspace{0.5cm}
\small
\item[*] NVIDIA GeForce GTX Titan X (Maxwell generation) characteristics
\end{tablenotes}
\end{threeparttable}
\label{tab:defs}
\end{table*}
\end{savenotes}

The parallel framework of choice for the accelerators is OpenCL, because it is vendor independent. In this regard GPU threads are referred as work-items, and GPU blocks of related threads are referred as work-groups. In AMBER, OpenCL kernels operate in three dimensional grids, but the pipeline uses only two dimensions, time and DM. These two dimensions limit the amount of available parallelism on both work-groups and work-items. The pipeline configuration is based on survey constraints and processing capabilities (see Table~\ref{tab:defs}) as well as 8 different types of user-controlled parameters\footnote{Previous pipeline version additionally had one more parameter \texttt{splitSeconds} responsible for manipulation between different kernels.} (see Table~\ref{tab:free}) that altogether define a single computational configuration for a specific many-core accelerator\footnote{Other input files contain downsampling factors, GPU cache line size, and frequency channels that need to be zapped due to terrestrial radio frequency interference (RFI) contamination.}.

\begin{savenotes}
\begin{table*}[t!]
\renewcommand*{\arraystretch}{1.5}
\centering
\caption{Types of user-controlled tuning parameters and their boundary conditions.}
\begin{threeparttable}
\scalebox{0.6}{
\begin{tabular}{c c c}
\hline\hline
\textbf{Parameter $p$} & \textbf{Description} & \textbf{Boundary conditions} \\
\hline
\texttt{localMem} & Utilization of local memory to allow for data re-use between computations & $p=0$ (no re-use); $p=1$ (total re-use) \\
\multirow{2}{*}{\texttt{unroll}} & \multirow{2}{*}{Loop unrolling to optimize code execution by means of its reorganization} & $p=1\dots\texttt{MAX\_UNROLL}$; \\
& & $\frac{\texttt{CHANNELS}}{\texttt{SUBBANDS}} \div p$ (step 1 de-dispersion), $\texttt{SUBBANDS} \div p$ (step 2 de-dispersion) \\
\multirow{2}{*}{\texttt{nrSamplesPerThread}} & \multirow{2}{*}{\textnumero\,samples per work-item} & $p = 1\dots\texttt{MAX\_ITEMS\_DIM0}$; \\
& & $\texttt{SAMPLES} \div p$ (both steps of de-dispersion) \\
\multirow{2}{*}{\texttt{nrDMsPerThread}} & \multirow{2}{*}{\textnumero\,DMs per work-item} & $p=1\dots\texttt{MAX\_ITEMS\_DIM1}$; \\
& & $\texttt{SUBBANDING\_DMs} \div p$ (step 1 de-dispersion), $\texttt{DMs} \div p$ (step 2 de-dispersion) \\
\hline
\multicolumn{3}{c}{$\texttt{nrSamplesPerThread} \times \texttt{nrDMsPerThread} < \texttt{MAX\_ITEMS}$ is regulated by the maximum number of registers on a GPU card} \\
\hline
\multirow{2}{*}{\texttt{nrSamplesPerBlock}} & \multirow{2}{*}{\textnumero\,samples per work-group} & $p=1\dots\texttt{MAX\_DIM0}$; \\
& & $\frac{\texttt{SAMPLES}}{\texttt{nrSamplesPerThread}} \div p$ (both steps of de-dispersion) \\
\multirow{2}{*}{\texttt{nrDMsPerBlock}} & \multirow{2}{*}{\textnumero\,DMs per work-group} & $p=1\dots\texttt{MAX\_DIM1}$; \\
& & $\frac{\texttt{SUBBANDING\_DMs}}{\texttt{nrDMsPerThread}} \div p$ (step 1 de-dispersion), $\frac{\texttt{DMs}}{\texttt{nrDMsPerThread}} \div p$ (step 2 de-dispersion) \\
\hline
\multicolumn{3}{c}{$\texttt{nrSamplesPerBlock} \times \texttt{nrDMsPerBlock} < \texttt{MAX\_THREADS}$ is regulated by the maximum number of OpenCL work-items} \\
\hline
\multirow{2}{*}{\texttt{nrItemsD0}} & \multirow{2}{*}{\textnumero\,items to process per work-item} & $p=1\dots\texttt{MAX\_ITEMS}$; \\
& & $\texttt{NRSAMPLES} \div p$ (time series downsampling); also $\texttt{SAMPLES} \div p$ (S/N calculation) \\
\multirow{2}{*}{\texttt{nrThreadsD0}} & \multirow{2}{*}{\textnumero\,work-items in a work-group} & $p=1\dots\texttt{MAX\_THREADS}$; \\
& & $\frac{\texttt{NRSAMPLES}}{\texttt{nrItemsD0}} \div p$ (time series downsampling); also $\frac{\texttt{SAMPLES}}{\texttt{nrItemsD0}} \div p$ (S/N calculation) \\
\end{tabular}
}
\end{threeparttable}
\label{tab:free}
\end{table*}
\end{savenotes}

All user-controlled tuning parameters need to be generated within the corresponding boundary conditions. For de-dispersion kernels, AMBER may or may not utilize local memory (\texttt{localMem}) and loop unrolling (\texttt{unroll}) to speed up the computations. The latter also scales with the \klmr{number} of channels distributed over the frequency subbands as this gives the amount of parallelism during the frequency channels summation. The \klmr{number} of time samples and DMs that each GPU has to correct for (\texttt{nrSamplesPerThread}, \texttt{nrDMsPerThread}, \texttt{nrSamplesPerBlock}, \texttt{nrDMsPerBlock}) are constrained by the overall number of time samples and DMs / subbanding DMs in the search space. Similarly, smoothing (S/N evaluation) kernels are computationally limited by the total (downsampled) number of time samples in the observations.

Before the real-time search, the pipeline undergoes brute-force (BF) auto-tuning to optimize each kernel by going through all possible configurations of tunable parameters from the single kernel parameter space. An average run of the BF tuning takes from 5 to 24 hours depending on the processing cluster architecture. At Westerbork, AMBER runs on the ARTS GPU Cluster~\citep{vanLeeuwen-2014}. \klm{For our tests, we only used the initial node of that cluster, ARTS0, powered by NVIDIA GeForce GTX Titan X GPUs.} On a single such GPU, the BF tuning takes about 10 hours. The runtime of the whole pipeline (all kernels) \klm{on the randomly generated test data\footnote{The data amount depends on the number of processed batches. In our experiments, we used 10 batches, each of 1.024\,sec, so the observation time is 10.24\,sec.}} with the tuned configuration lasts about $T_\mathrm{BF} \simeq 5.5$\,sec. 

The complete sampling has several drawbacks:
\begin{itemize}
\item \klmr{The BF tuning does not tune all kernels at once as it takes too much time, even on a large parallel system (see Section~\ref{sec:discussion});}
\vspace{0.25cm}
\item \klmr{As the BF tuning is applied to each kernel separately, it does not consider dependencies between kernels which may lead to a global optimal configuration for the whole pipeline.}
\end{itemize} 

In this paper we test the idea that more heuristic genetic algorithms can find a \klm{good enough} global configuration for all kernels together \klm{in less time}. Although the end configuration may not be the overall global optimum (see Section~\ref{sec:discussion}), it will still be nearly as good or even better than the combination of best configurations per kernel after BF tuning. Additionally, it will almost always be better than the best configuration after a pure random search. The idea and overview of the GA algorithm is given in the following section. 

\section{The genetic algorithm}\label{sec:ga}
The idea of genetic algorithms (GAs) dates back to Charles Darwin's idea of biological evolution that only the fittest individuals should survive and produce more adapted offspring~\citep{Holland-1975}. Most applications for GAs are in search and optimization~\citep{Goldberg-1989}. There are also a number of GA applications in astronomy, from spectral analysis and cosmology to telescope scheduling~\citep{Charbonneau-1995, Metcalfe-2000, Mokiem-2005, Liesenborgs-2006}, but also in searches for pulsars and gravitational waves~\citep{Lazio-1997, Petiteau-2013}. 
\subsection{Main genetic operators}
\begin{figure}[t!]
\centering
\includegraphics[width=0.5\linewidth]{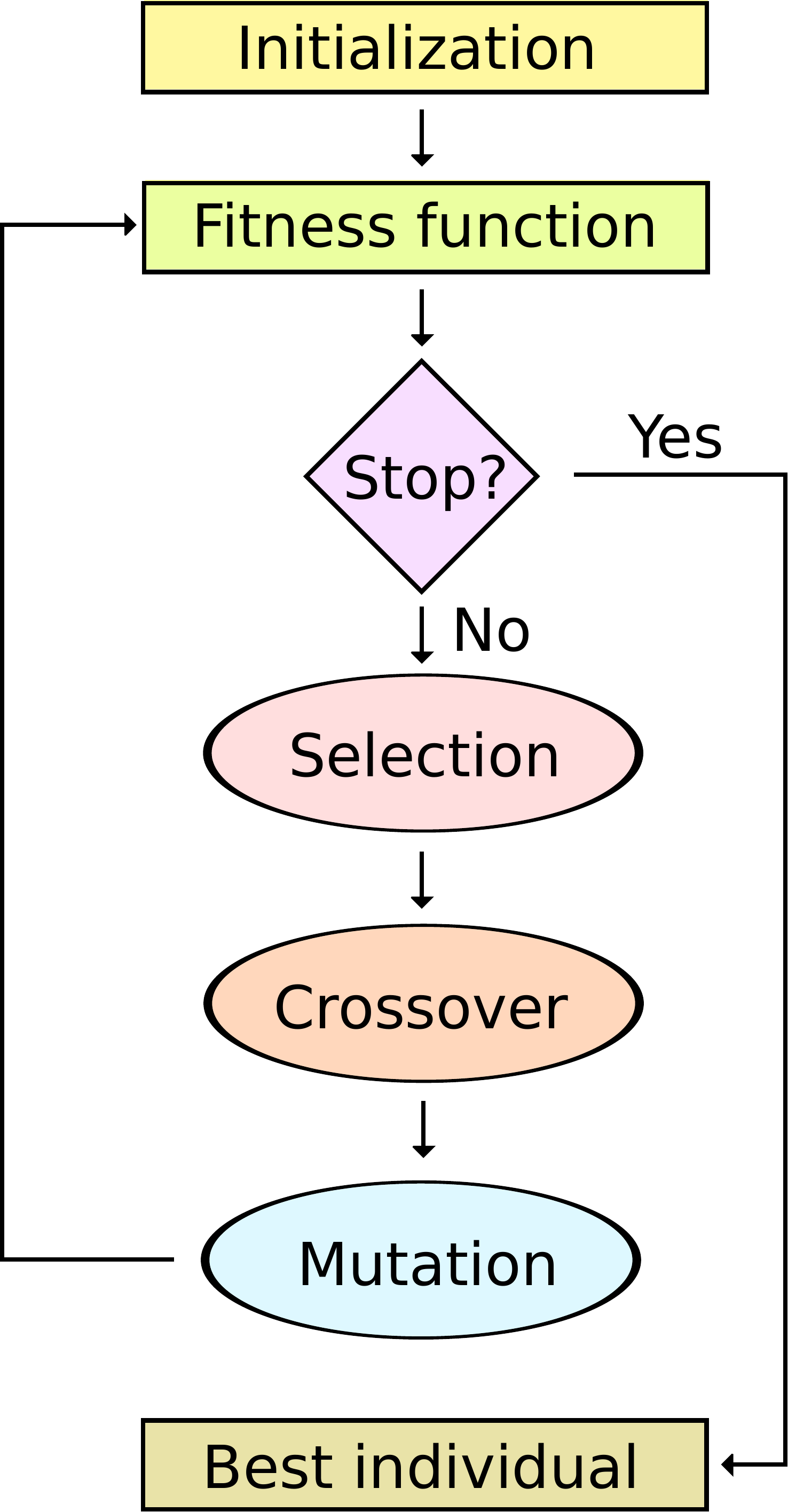}
\caption{Typical block diagram of a genetic algorithm. After initializing population of individuals, we evaluate fitness functions of each individual and check the stopping criteria. We either finish or make individual selection based on its fitness value. Next, we apply crossover and mutation to selected individuals, re-evaluate their fitness functions and check the stopping criteria again. Once we satisfy the criteria, we end up with the best individual from the evolved population. Otherwise we replace old individuals by new ones and continue evolution further.}
\label{fig:ga}
\end{figure}

The working element of every GA is a chromosome or an individual -- a set of tunable (usually binary) parameters known as genes. The idea of GA is to evolve chromosomes and improve their scores guided by a fitness function. A typical GA contains five main genetic operators: initialization, selection, crossover, mutation, and replacement (see Fig.~\ref{fig:ga}).

The individuals are first initialized (usually at random) and acquire their respective fitness functions. Based on the fitness scores, a number of fittest individuals get selected for further evolution. The selection is typically done either via roulette-wheel scheme (based on cumulative probability of fitness functions) or tournament scheme (based on the fittest winner from a limited pool of individuals). 

The pair of selected individuals then produces an offspring through mixing their genes among each other. Such operation is known as crossover. The most used option to cross genes is a one-point crossover where the genes before the crossing point remain the same, but the rest of the genes gets swapped with that from another individual. Depending on the type of genes and the task, other crossover options such as multi-point or uniform crossover are possible~\citep[see][for a review]{Umbarkar-2015}. 

Next, offspring individuals may also undergo mutation when one or more genes happen to randomly change their values~\citep[e.g.,][]{Soni-2014}. This is done primarily to avoid premature convergence into a local optimum and better explore parameter space. 

Finally, the new population replaces the old one with possible preservation of fittest parent individuals. After that the evolution starts again until the population obtains sufficient fitness function or the evolution reaches its limit in time or in the number of generations. The size of the individuals population $N_\mathrm{pop}$ as well as the rates of crossover $P_\mathrm{cross}$ and mutation $p_\mathrm{mut}$ are tunable parameters of the GA algorithm.
\subsection{GA auto-tuning}
The main advantages of the GA approach compared to BF tuning or random search during AMBER auto-tuning optimization are:
\begin{itemize}
\item The BF algorithm tunes each kernel separately, whereas GA tunes the whole pipeline and thus considers interactions between the kernels. As a result, \klm{GA can find a better solution for all kernels together in less time}.
\vspace{0.25cm}
\item Unlike random search, GA does a guided search for better parameters while still trying to explore the rest of the parameter space by means of mutation operator. 
\end{itemize}

In our GA implementation\footnote{The code is available on GitHub:~\url{https://github.com/MixKlim/GA_AMBER}}, every individual is a set of free parameters \klm{(see Table~\ref{tab:free})} that altogether control all four kernels of the pipeline. Only \texttt{localMem} gene has a binary representation and switches between `0' and `1', other genes are powers of two for simplicity, and to preserve bit alignment on GPUs. 

We first generate arrays of possible values for every gene based on boundary conditions and dependencies (see Section~\ref{sec:amber}). Next, all four kernels of $N_\mathrm{pop}$ individuals obtain genes initialized with random values from those arrays. After that we evaluate individuals fitness functions \klm{as the run times of the whole pipeline, which includes a combination of all four kernels.} We then apply tournament selection by creating a pool of randomly selected individuals $N_\mathrm{pool}$ \klm{(some fraction of the total population $N_\mathrm{pop}$)} such that the individual with the best fitness wins and gets chosen from such pool. We do selection $N_\mathrm{pop}$ times.

\begin{figure}[t!]
\centering
\includegraphics[width=0.8\linewidth]{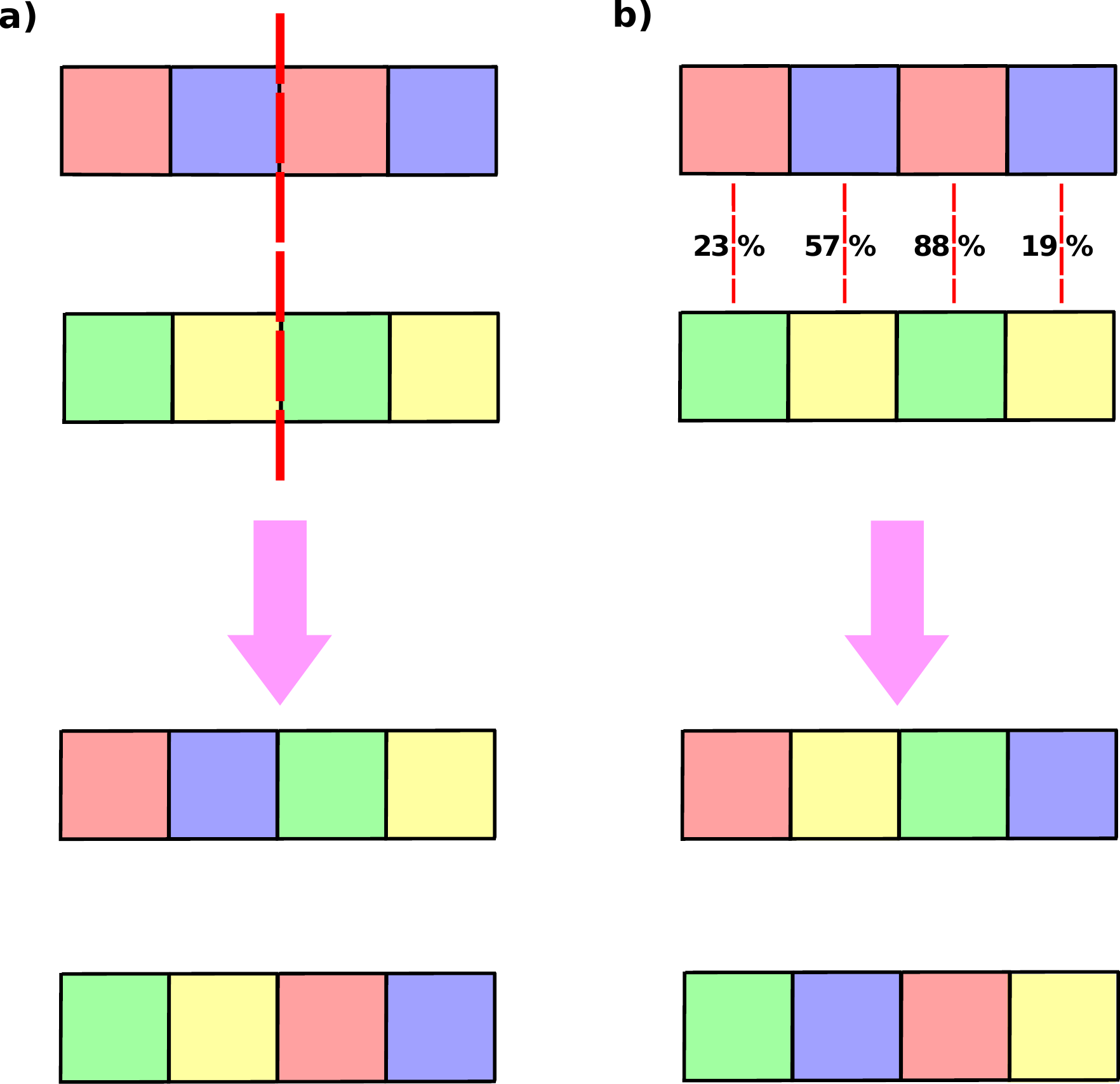}
\caption{Two types of crossover that we test in our GA: a) one-point crossover; only the genes (coloured squares) after the red dashed line get exchanged among two individuals. The line crossing point is chosen at random; b) uniform crossover with a coin toss probability; the exchange of genes between two individuals happens only if the randomly generated probability associated with these genes exceeds 50\%.}
\label{fig:cross}
\end{figure} 

Next we group selected individuals in pairs for the subsequent crossover. We make sure there are minimal or no identical individuals in pairs as this leads to no new offspring. After that we apply crossover for each kernel with probability $p_\mathrm{cross}$ to create an offspring population of individuals with interchanged genes. The size of offspring population is the same as the size of their parents $N_\mathrm{pop}$. We test two types of crossover: one-point crossover and uniform crossover with a coin toss probability (see Fig.~\ref{fig:cross}). The gene exchange happens only if the offspring genes also satisfy required boundary conditions.

Some kernels of each offspring individual then also undergo mutation with probability $p_\mathrm{mut}$; one or multiple genes in these kernels get randomly changed from an array of its possible values. For kernels that include downsampling (smoothing and S/N kernels), a number of genes get randomly chosen for mutation, one per downsampling factor. We make sure the new gene value is different from the old one unless it is prohibited by the boundary conditions. After mutation all offspring individuals get their fitness functions re-evaluated. 

In the end, we replace parent individuals by their offspring. \klm{We rank parent and offspring individuals by their fitness values \klm{and select the best performing half of each group}. In this case we preserve the fittest parent individuals and have a new population of the same size $N_\mathrm{pop}$ for the next generation.}

\section{Performance results}\label{sec:results}
Since we are interested in testing \klm{how much faster GA tuning finds a solution nearly as good as BF tuning}, we run our genetic evolution for as long as it takes BF to explore and tune every single kernel, i.e. $T_\mathrm{run} \simeq 10$\,hrs for ARTS0. To evaluate fitness values we run AMBER with each individual configuration on \klm{the uniformly distributed noise with injected single pulse signal} and obtain a set of execution times $T_\mathrm{exec}$. Although the total runtime $T_\mathrm{total} = T_\mathrm{data} + T_\mathrm{exec}$ also includes time spent on test data generation $T_\mathrm{data}$, we do not take that time into account while evaluating the individual's fitness function. Nevertheless, some configurations can result in a very slow run of the pipeline or even its breakdown, mostly due to inappropriate memory allocation. To avoid such configurations, we limit the AMBER total runtime $T_\mathrm{total}$ to 3 min. Such \klm{empirical time limit was chosen to cover half-minute fluctuations from test data generation and most typical pipeline executions, but also penalize inefficient runs.} For the tournament selection, the size of the tournament pool ${N_\mathrm{pool}}$ was set to be $20\%$ of $N_\mathrm{pop}$ to avoid multiple selections of only several dominant individuals.

In our tests we used $N_\mathrm{pop} = 20$ individual configurations to balance between slow fitness function evaluation and sufficient number of generations. Unless being tested, crossover and mutation were applied with $p_\mathrm{cross} = 0.8$ and $p_\mathrm{mut} = 0.1$, somewhat generally accepted average rates in the literature~\citep[see][for a review]{Patil-2015}. We also recorded configurations with the best fitness value after each generation. We then tracked the best fitness value in the population over all generations along with the computational time spent per generation. 

\begin{figure*}[t!]
\centering
\subfigure[Evolution with different $p_\mathrm{mut}$ probability.]{
\includegraphics[width=0.48\textwidth, trim={0.6cm 0.1cm 1.0cm 0.9cm}, clip]{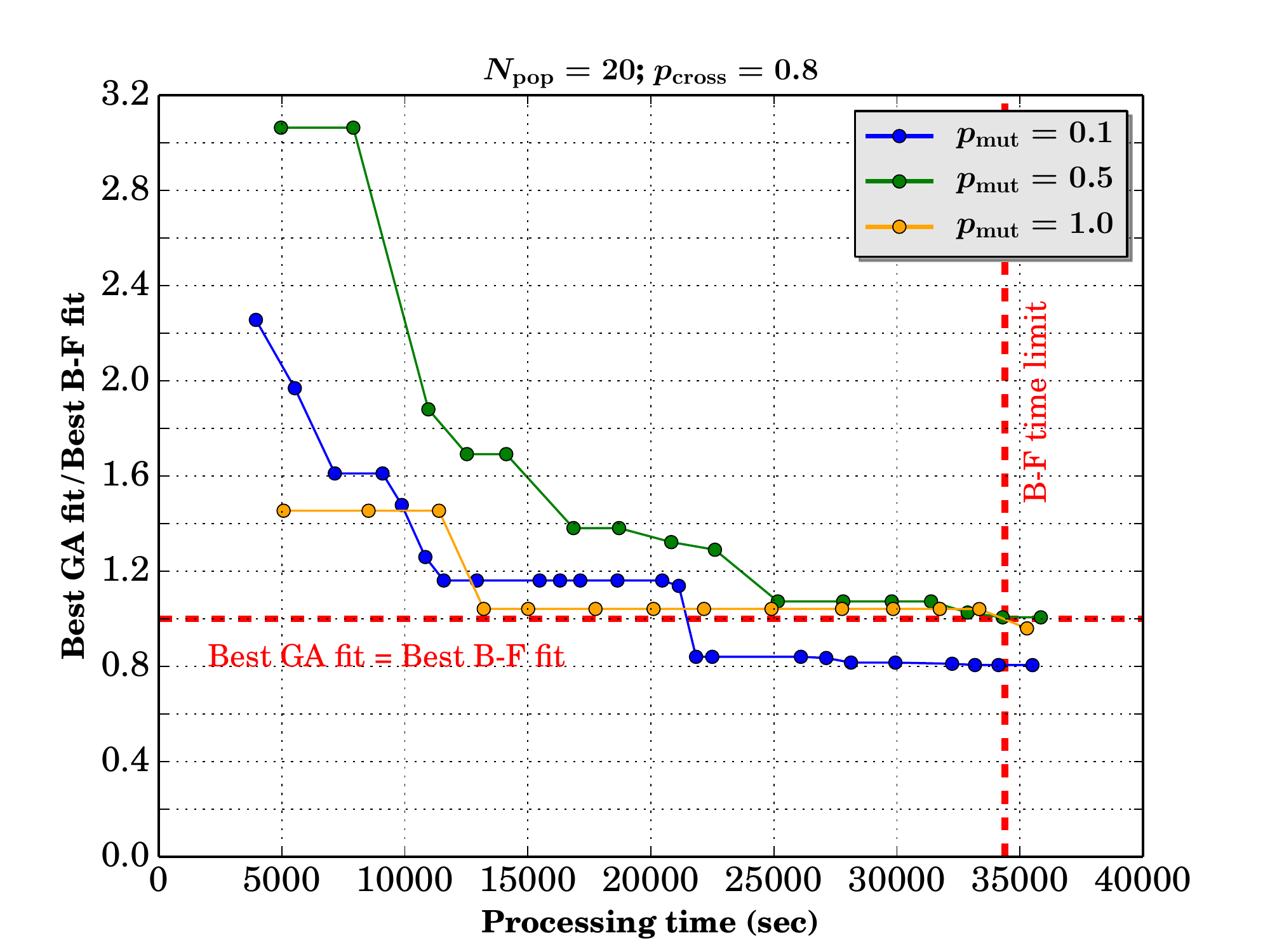}\label{fig:muttest}
}
\subfigure[Evolution with different $p_\mathrm{cross}$ probability.]{
\includegraphics[width=0.48\textwidth, trim={0.6cm 0.1cm 1.0cm 0.9cm}, clip]{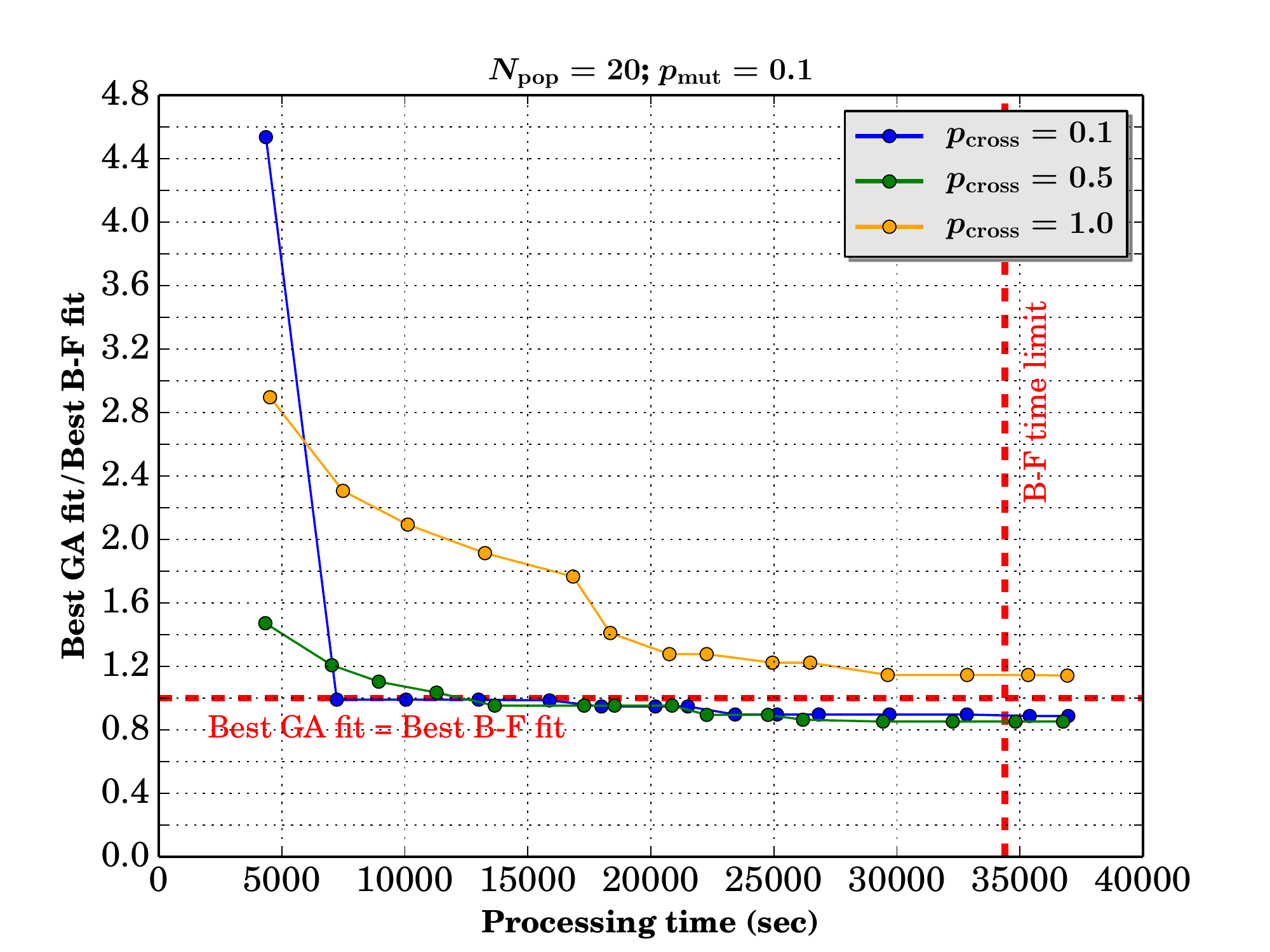}\label{fig:crosstest}
}
\subfigure[Evolution with different $N_\mathrm{pop}$ size.]{
\includegraphics[width=0.48\textwidth, trim={0.6cm 0.1cm 1.0cm 0.9cm}, clip]{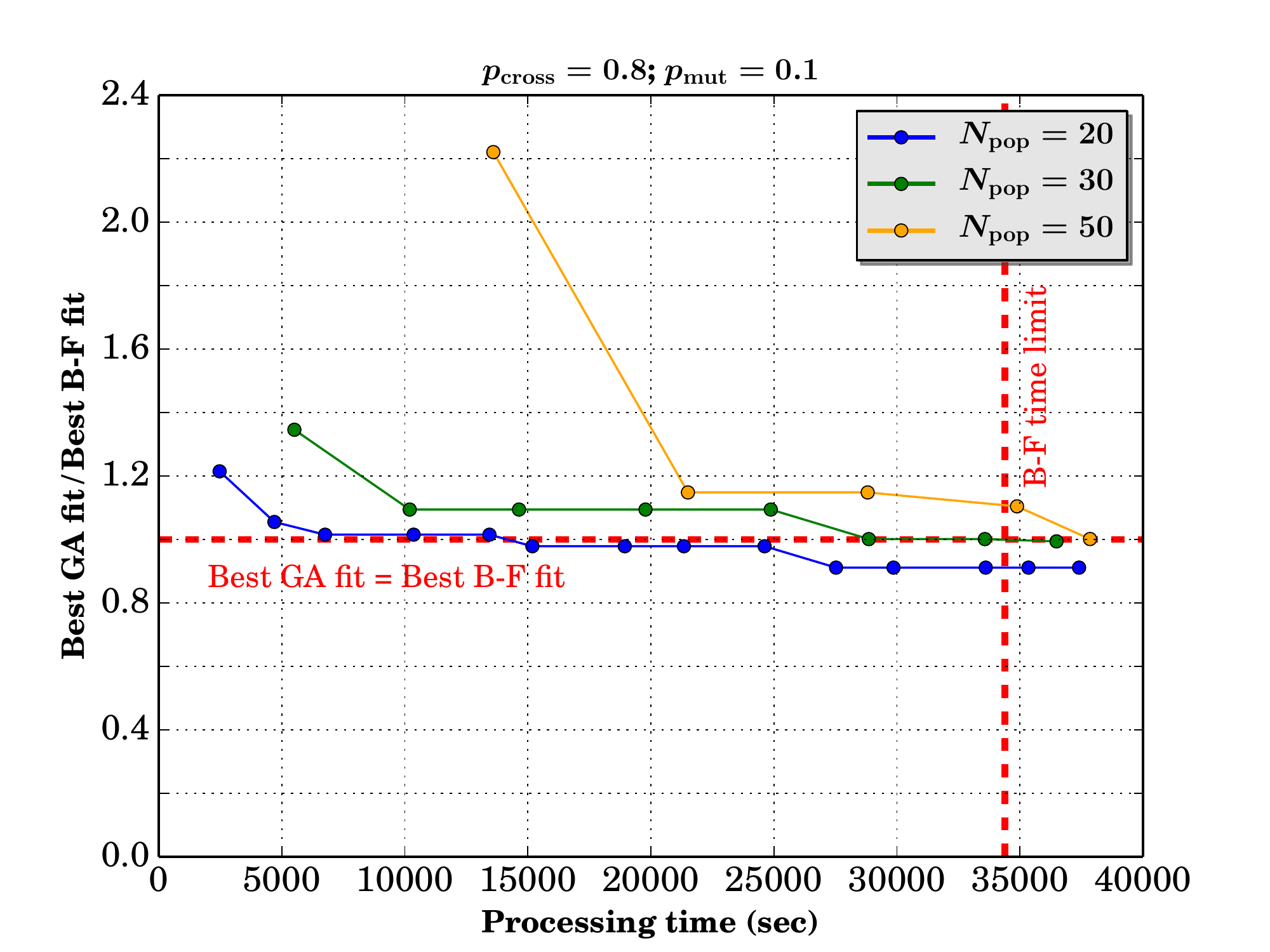}\label{fig:sizetest}
}
\subfigure[Different types of crossover \& random search option.]{
\includegraphics[width=0.48\textwidth, trim={0.6cm 0.1cm 1.0cm 0.9cm}, clip]{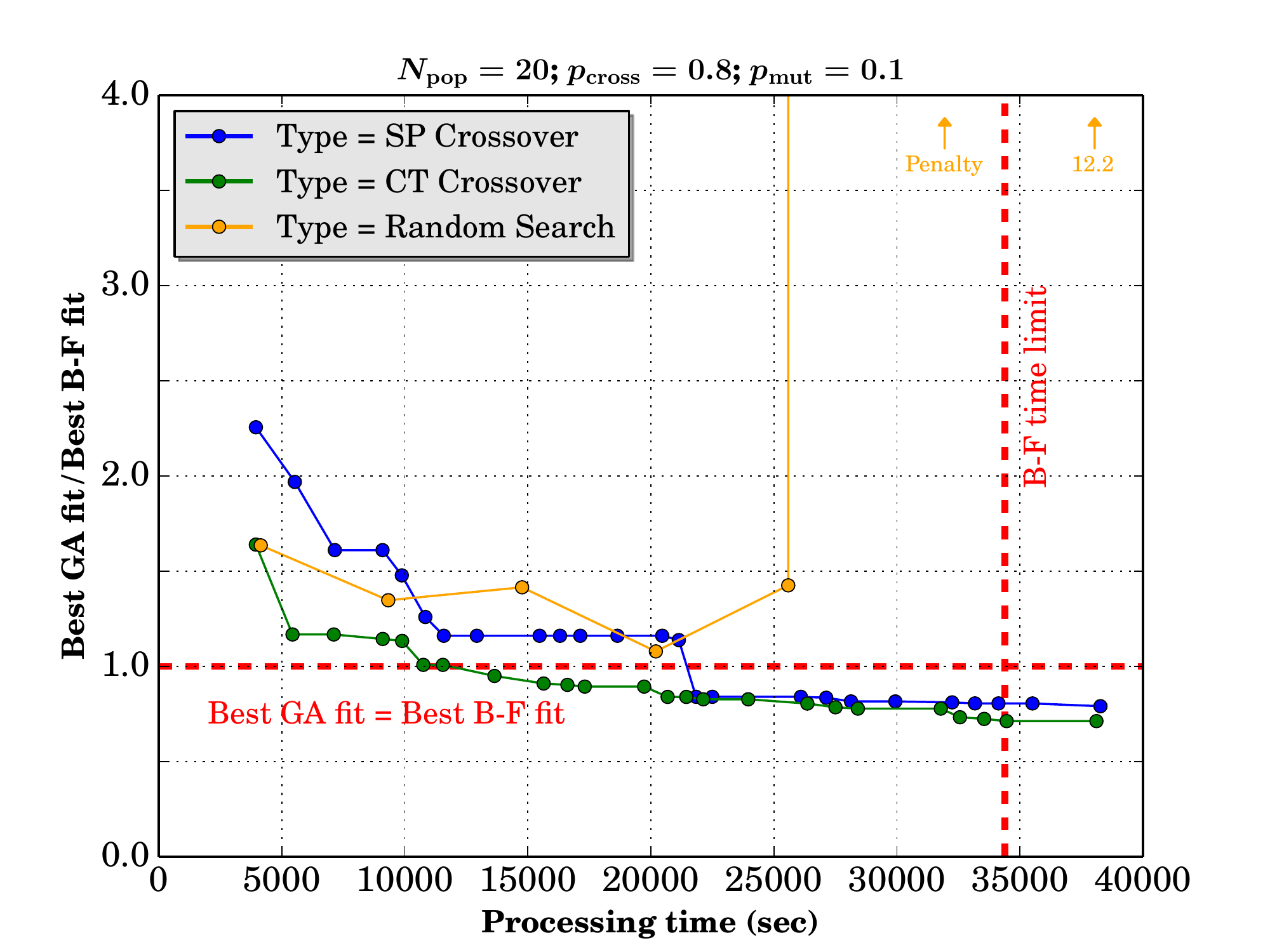}\label{fig:typetest}
}
\caption{Different tests on algorithm convergence with (\ref{fig:muttest}) changing mutation rate; (\ref{fig:crosstest}) changing crossover rate; (\ref{fig:sizetest}) changing population size; (\ref{fig:typetest}) changing type of crossover (single-point, toss-coin) and performing pure random search. The last two runs of the random search got overall bad or penalized fitness values that appear higher and are therefore not visible on the plot.}
\label{fig:tests}
\end{figure*}

We did several 10\,hr test runs based on different population size, probabilities of crossover and mutation, type of crossover operator, and pure random search. For each run and for every population generation, we plot the ratio between the best obtained execution time over the runtime found by BF tuning $T_\mathrm{exec} / T_\mathrm{BF}$. \klmr{The algorithm performance plots for every aforementioned parameter are shown in Fig.~\ref{fig:tests}; each data point is based on a single test run, since the algorithm includes a random component. The gradual fitness improvement, however, does not change with different runs (in other words, the evolution does not diverge).}

In case of changing mutation rate (Fig.~\ref{fig:muttest}), more frequent mutation leads to a more complete exploration of the total parameter space. Still, even when individuals always happen to mutate during their evolution ($p_\mathrm{mut} = 1$, best exploration), it may take much longer time compared to BF runtime to significantly improve the best configuration in population, mostly due to high complexity of the parameter space we have to explore (see Section~\ref{sec:discussion}) and prolonged fitness evaluations. 

In case of changing crossover rate (Fig.~\ref{fig:crosstest}), more frequent crossover \klm{generally leads to a better mixing between individuals and therefore more rapid population evolution}. Depending on how good the best configuration gets after random initialization, higher crossover rates result in higher evolutionary slopes. Extremely high rates ($p_\mathrm{cross} = 1$, all individuals get updated and then mutated) do not necessarily lead to the best fitness at the end of GA evolution, but greatly reduce its value after initialization. 

The more individuals we initialize, the more time we spend on generating arrays of possible values for each gene based on boundary conditions (see Section~\ref{sec:amber},~\ref{sec:ga}). This is illustrated in Fig.~\ref{fig:sizetest}, where we test GA performance relative to the population size. As a result, \klm{GAs with larger populations undergo less iterations within $T_\mathrm{run}$ and are thus less likely to converge into good solutions.} On the other hand, larger populations can better sample total parameter space during initialization, which can still result in a good fit (see also~\cref{app:appendix}). 

Fig.~\ref{fig:typetest} shows tests based on different types of crossover operator, single-point and toss-flip (see also Fig.~\ref{fig:cross}). Although toss-flip crossover implies more gene exchanges than single-point crossover, this does not reflect on the overall convergence of the best fit. We also test pure random search option where we do not use genetic operators but randomly initialize new populations until we reach $T_\mathrm{run}$. In this case there is no interaction between individuals or evolution of their population. Again, initialization takes more time than evolution, therefore we end up with less amount of runs before we reach BF time limit. Also, we do not track best individual configurations during $T_\mathrm{run}$ and thus can have penalties in fitness functions for certain random search runs. Nevertheless, we may sometimes get relatively good fits straight from the initialization but those are rare due to the complexity of the total parameter space (see Section~\ref{sec:discussion}).

All plots show that the GA evolution can be generally described by a rapid drop of fitness value at the early evolution stage (subject of initialization and selection) and its much slower improvement at the later evolution stage (subject of crossover and mutation). Thus, we can already find a reasonably good configuration of $1.5-2\times T_\mathrm{BF}$ after $2-5$ hours of GA evolution. We also see that within $T_\mathrm{run}$ GA almost always converges into a better solution for the whole pipeline than BF tuning for each kernel. Again, this is because GA fitness function is guided by the \klm{overall} performance of the whole pipeline, whereas BF optimizes each kernel separately. \klm{None of the GA parameters drastically change the gradual fitness improvement and its proximity to the best BF solution.} The later evolution of the search parameters happens very slow and is also quite independent of the population size. As the algorithm converges to one of the local, good configurations after initialization or during very first generations, it can still improve them later through crossover or even find the best, global configuration through mutation.

\section{Discussion}\label{sec:discussion}
The main downside of every fitting algorithm is that after finding a local, degenerate solution, it is very unlikely to improve and converge into a much better, global fit. Our GA performance tests show that the parameter space we are trying to fit is very complex and possesses many local configurations that give almost identically good but not necessarily the best performance. To check the variety of parameter configurations that GA converges to, we build up histograms for best individual's genes evolved in populations of three different sizes: $N_\mathrm{pop} = 20$, $30$, and $50$ (see~\cref{app:appendix}). The diversity in explored parameter ranges as well as histogram shapes shows that we obtain multiple degenerate solutions at the end of each GA evolution. Therefore, it takes more time than $T_\mathrm{run}$ for one single run of GA evolution to cover all good configurations and determine the best among them.

\begin{figure}[t!]
\centering
\includegraphics[width=\linewidth, trim={0cm 0.25cm 0cm 0cm}, clip]{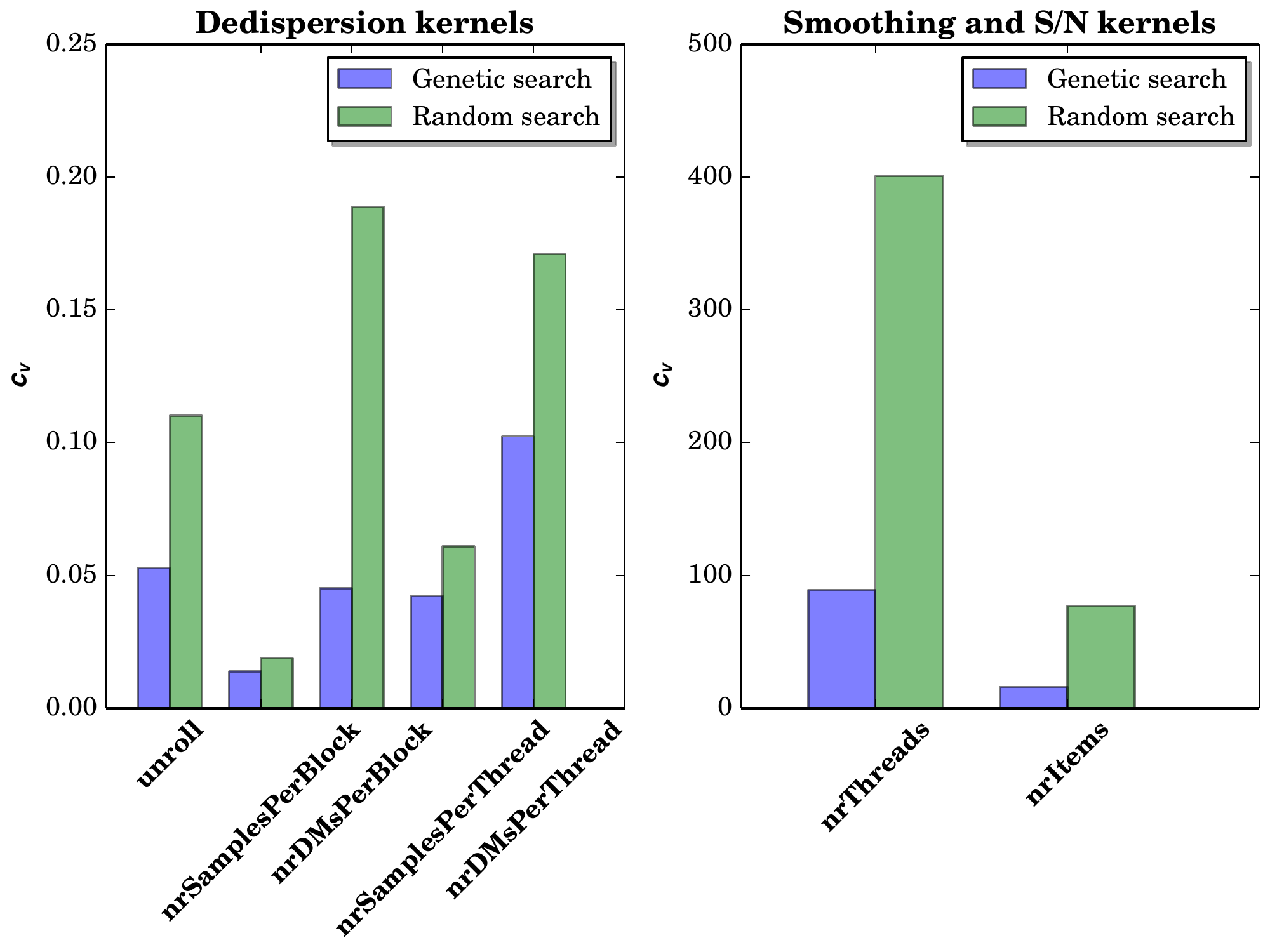}
\caption{Bar histogram of coefficients of variation $c_v$ among $N_\mathrm{pop} = 20$ individual configurations averaged over 30 GA and random searches. GA search has overall smaller $c_v$ than random search as it utilizes genetic memory which gives parameter convergence. \texttt{nrThreads} and \texttt{nrItems} get high $c_v$ since they have more freedom to change due to multiple downsampling factors. The coefficient of binary \texttt{localMem} variation is not present as it approaches infinite values once there is only a small variation in different configurations.}
\label{fig:cov}
\end{figure}

To test the algorithm convergence, we measure the coefficient of variation $c_v = \sigma / \mu$ among different configurations at the end of every algorithm run; $\sigma$ is the standard deviation of the given parameter and $\mu$ is its mean value in a set of end configurations. We then average $c_v$ over multiple algorithm runs. The gaps in parameter ranges caused by boundary conditions do not affect $c_v$ as both $\mu$ and $\sigma$ get affected but balance each other in a ratio. Higher $c_v$ shows more diversity in individual genes, whereas the algorithm convergence requires low $c_v$. Fig.~\ref{fig:cov} shows averaged coefficients of variation among 20 configurations after 30 GA and random search executions. We see that GA shows strong parameter convergence compared to memoryless random search, and thus results in smaller $c_v$. Most diversity happens in smoothing and S/N kernels since with multiple downsampling factors, these kernels have more freedom to get their parameters changed. However, de-dispersion kernels represent the highest pipeline workload. As a result, both one-step and two-step de-dispersion contribute in a greater degree to the execution time\footnote{$30-70\%$ of $T_\mathrm{exec}$ based on an average runtime for $20$ configurations.}, and are thus most crucial for tuning. We do not treat the variation of the binary parameter \texttt{localMem} as it has a near-zero mean and is thus very sensitive to small diversities in various configurations.

To better see how well GA tunes the pipeline compared to BF search, we estimate the volume of the tuneable parameter space. Despite the fact we only have 8 different types of free parameters (Table~\ref{tab:free}), their number grows and thereby expands total parameter space as we consider multiple de-dispersion and donwsampling steps. Each de-dispersion kernel has six free parameters, whereas two other kernels, signal smoothing and S/N evaluation, have $2$ free parameters for every downsampling factor. Given $9$ downsampling factors for Apertif and taking additional S/N evaluation for a non-downsampled signal into account, we have in total $2\times6 + 9\times2 + 10\times2 = 50$ free parameters. As we initialize individual genes from arrays of possible values determined by boundary conditions, we can estimate how many possible values each individual gene can have. Since user-controlled parameter ranges are independent between different pipeline kernels, we get up to $2\times10^{15}$ possible configurations. Sampling that many configurations for the whole pipeline with a 3 min runtime limit would require $1.9\times10^8$ years, impossible \klm{even with a large parallel system}.

It is also hard to predict a global optimal configuration for a pipeline without knowing the landscape of such high-dimensional parameter space. Nevertheless, even though different genes from various best individuals do not resemble each other, it is the combination of all genes that determines individual fitness function, or the performance of the whole pipeline. Thus, if we are determined to get a reasonably good configuration in few hours, finding one local solution and evolving it is enough to reach a better \klm{overall} performance than applying a much deeper and longer BF search \klm{for each kernel}. Our tests show that for ARTS0 a combination of random search at the beginning and evolution later can in $2-5$ hours get just as good or even better configuration than what BF approach can obtain in $10$ hours.

\begin{figure}[t!]
\centering
\includegraphics[width=\linewidth, trim={1.0cm 0.3cm 1.0cm 1.0cm}, clip]{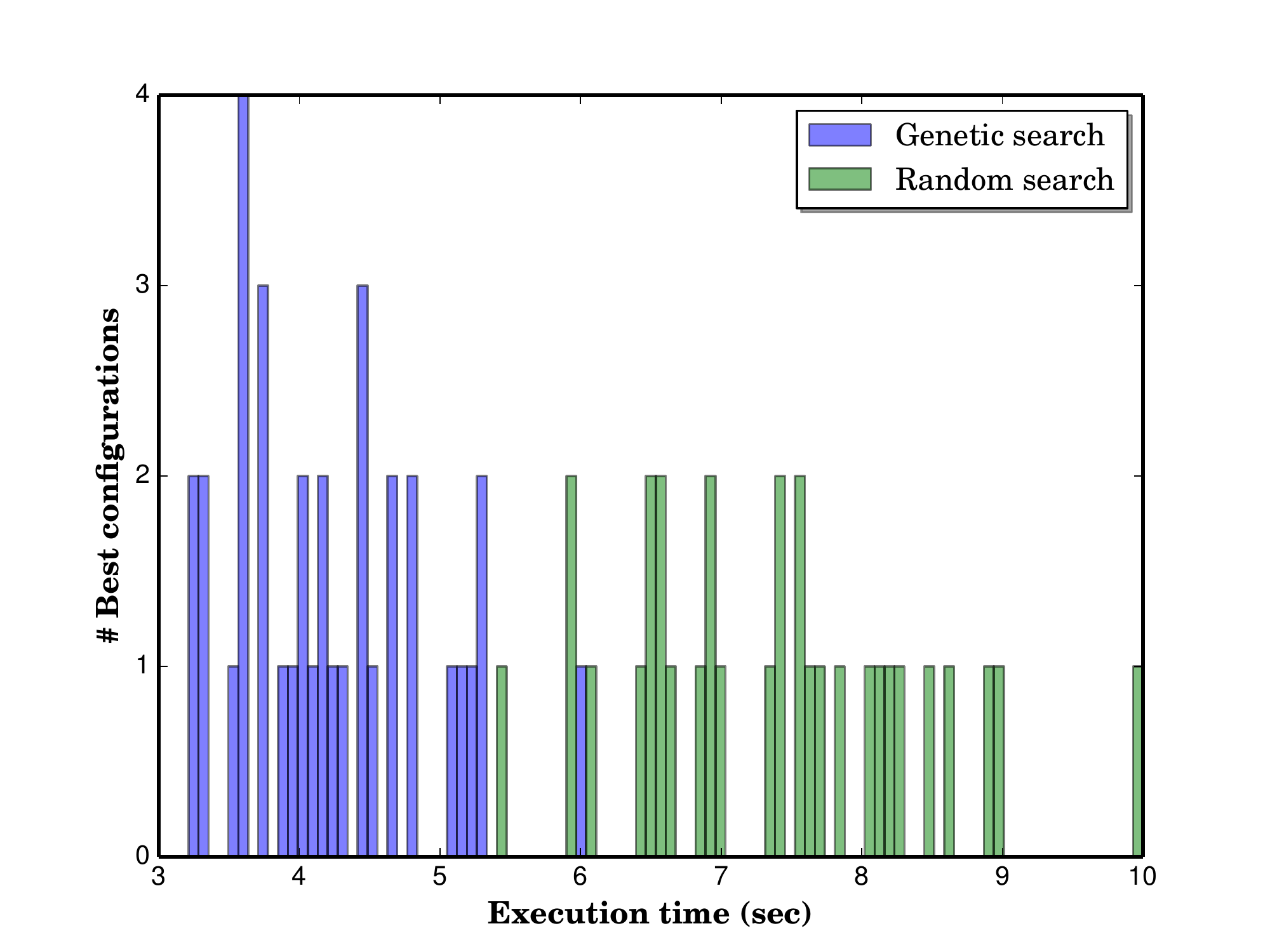}
\caption{Histogram of best fitness values after 35 GA and random searches. Although random initialization may sometimes reach good fits, GAs generally evolve their individuals to much better \klm{overall} performance. In general, it takes more time to initialize new individuals within boundary conditions rather than evolve already good solutions \klm{that are a priori within such conditions}.}
\label{fig:evalhist}
\end{figure}

This also raises the question whether a complete random draw of configurations from the complete parameter space (pure random search) would do just fine. The histogram of best fitness values based on GA evolution and complete random search is given in Fig.~\ref{fig:evalhist}. We see that on average GA finds better solution than just a random search, although the latter can sometimes overtake due to ``lucky shots''. However, as in every random process, there is no time certainty on how long we might have to wait before a reasonably good individual gets initialized. In GAs there is a constant fitness improvement that constraints the expected waiting time to $2-5$ hours for ARTS0 instead of $10$. Furthermore, good random picks should in general be rare as the total parameter space is very large-scale and hard to fit without any evolution.

\section{Conclusions}\label{sec:final}
Real-time single pulse searches are computationally intensive. Multiple factors play key role in how deep and fast searches can be made: total field of view and sensitivity in the survey, memory bandwidth and data rate of the beamformer, computational performance and versatility of the backend. We need sophisticated pipelines to speed the data processing up.

Before the actual processing, the transient search pipeline AMBER finds the most optimal configuration of user-controlled parameters for every pipeline kernel so that each kernel can perform at its fastest. This gets achieved via brute-force exploration of every kernel parameter space and takes many hours of processing depending on a survey setup. Besides, this does not take any dependencies between kernels into account and therefore does not tune the pipeline as a whole. Such tuning strategy does not allow the pipeline to be quickly retuned in case the pipeline gets modified or upgraded. 

Our search strategy based on genetic algorithms shows that with GAs we can always obtain a nearly as good or even better configuration for the whole pipeline in less amount of time, i.e. $2-5$ hours for ARTS0. The better the configuration gets obtained during the random initialization, the faster the GA converges into an already good fit. Apart from that, strong selection together with frequent crossovers and casual mutations will always handle badly initialized population and still lead to a good fit in the end. Such strategy can be easily ported to more sensitive pipelines and surveys.

Heuristic algorithms are a perfect tool to quickly obtain a local solution that can be nearly as good, or even better, than BF tuning of each kernel. For multidimensional parameter spaces in radio astronomy and other domains (bioinformatics, cryptography), heuristics is by far the easiest way to find reasonably good solution in a short period of time and with limited computational resources. 

\section*{Acknowledgements}
The development and commissioning of ARTS is carried out by a large team of engineers and astronomers, including active participation from the authors and PI (Joeri van Leeuwen). The research leading to these results received funding from the Netherlands Research School for Astronomy under grant NOVA4-ARTS (KM) and from the Netherlands eScience Center under grant AA-ALERT, 027.015.G09 (AS).

\section*{References}
\bibliography{GA_AMBER}{}
\bibliographystyle{elsarticle-harv}

\appendix
\label{appendix}
\gdef\thesection{\Alph{section}} 
\makeatletter
\renewcommand\@seccntformat[1]{Appendix \csname the#1\endcsname.\hspace{0.5em}}
\makeatother
\crefalias{section}{appsec}
\section{Evolutionary histograms for the best individuals in populations of different sizes}\label{app:appendix}
\begin{landscape}
\begin{figure}
\centering
\includegraphics[width=0.85\linewidth, trim={0.3cm 0.3cm 0.3cm 0.3cm}, clip]{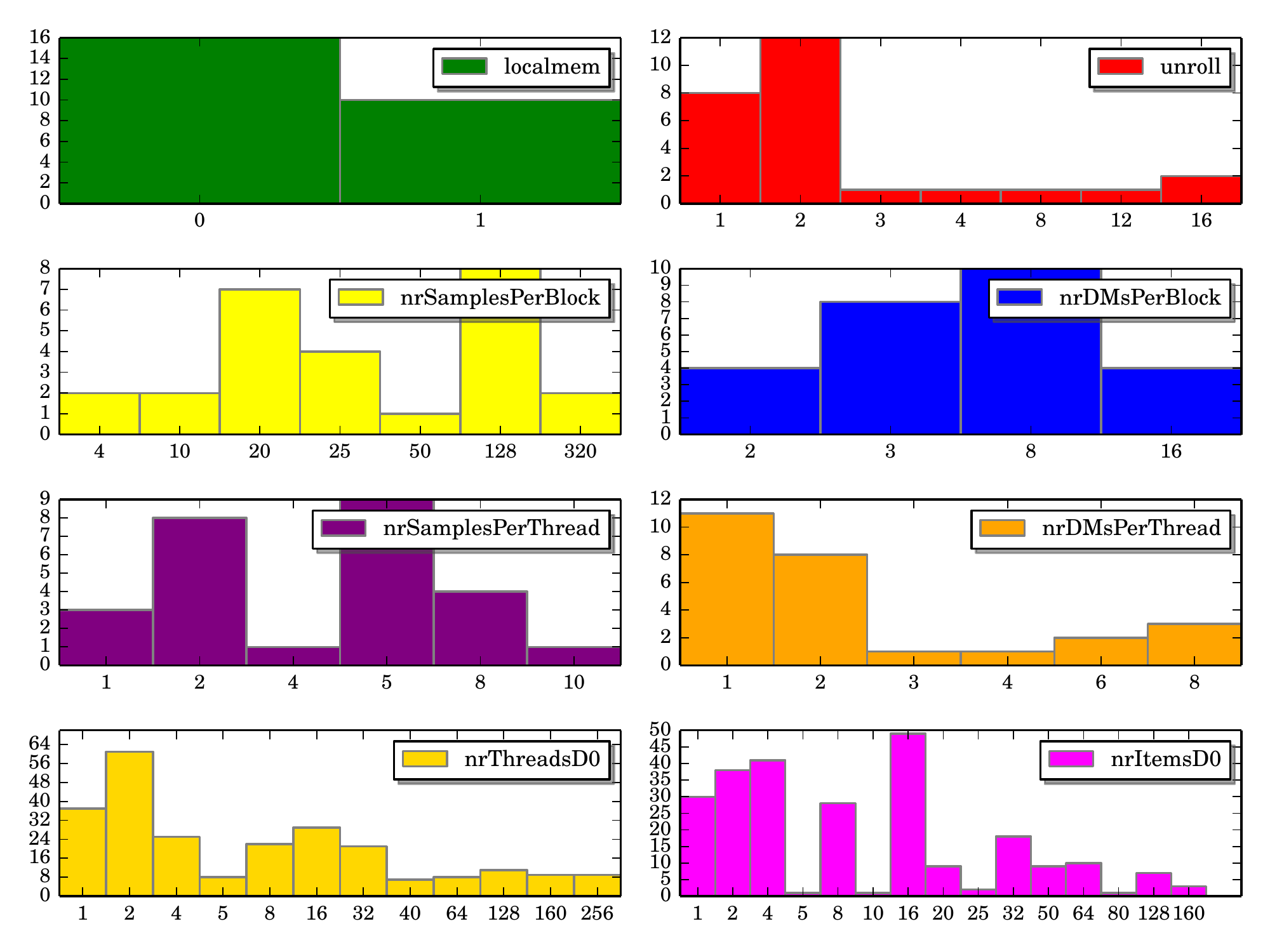}
\caption{Genetic evolution with $N_\mathrm{pop} = 20$ individuals.}
\label{fig:partest1}
\end{figure}
\end{landscape}

\begin{landscape}
\begin{figure}
\centering
\includegraphics[width=0.85\linewidth, trim={0.3cm 0.3cm 0.3cm 0.3cm}, clip]{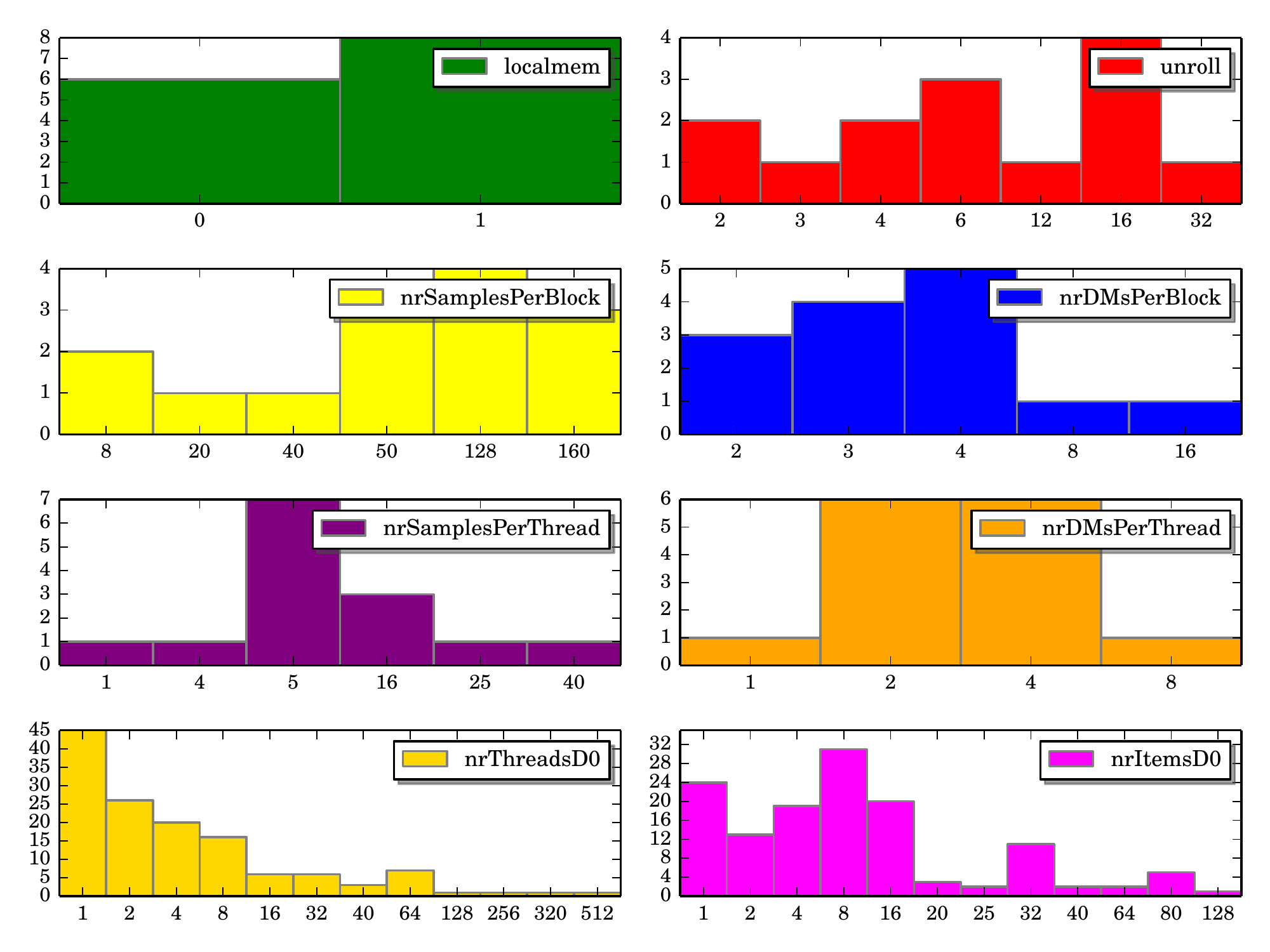}
\caption{Genetic evolution with $N_\mathrm{pop} = 30$ individuals.}
\label{fig:partest2}
\end{figure}
\end{landscape}

\begin{landscape}
\begin{figure}
\centering
\includegraphics[width=0.85\linewidth, trim={0.3cm 0.3cm 0.3cm 0.3cm}, clip]{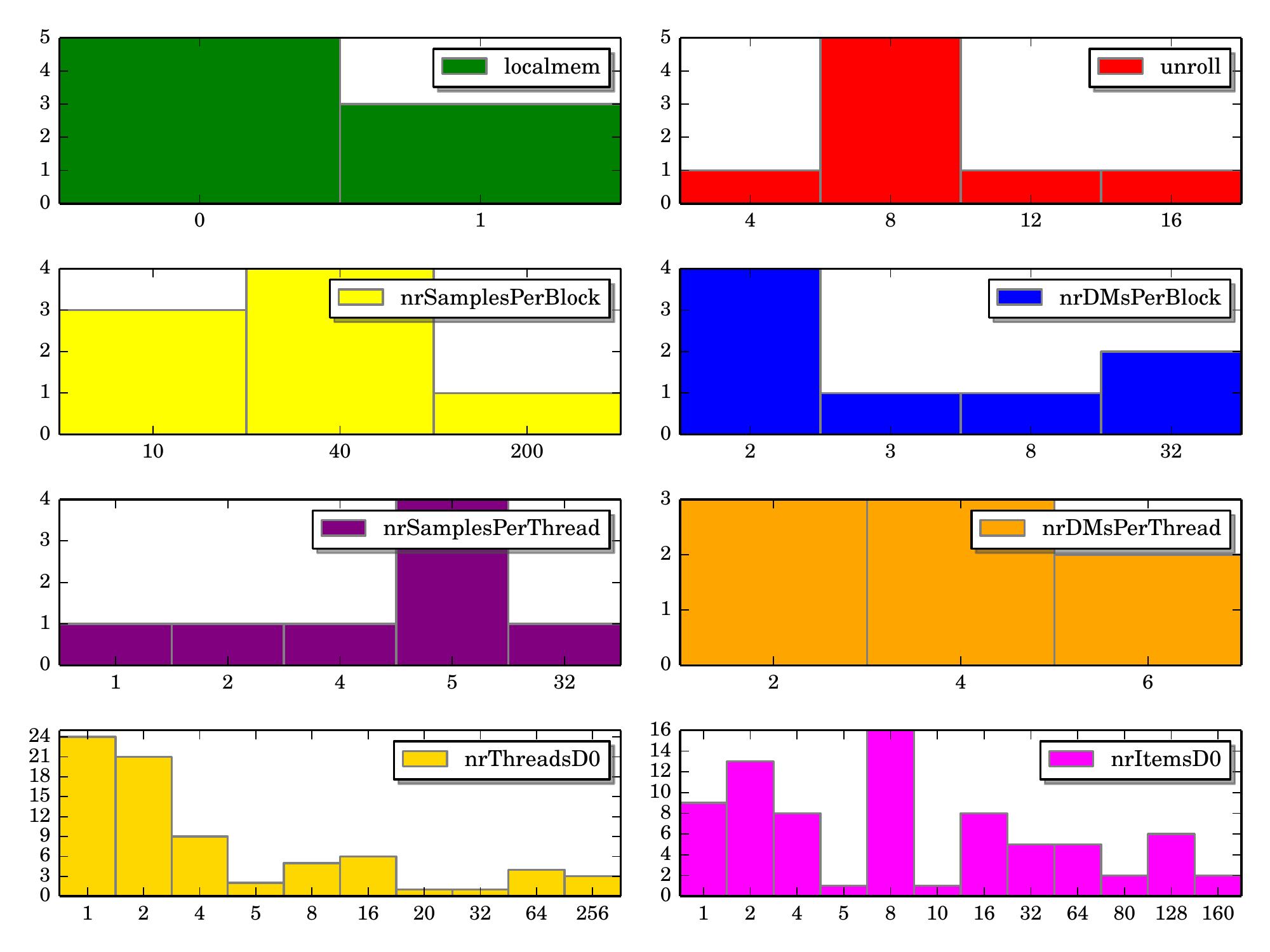}
\caption{Genetic evolution with $N_\mathrm{pop} = 50$ individuals.}
\label{fig:partest3}
\end{figure}
\end{landscape}

Figures~\ref{fig:partest1},~\ref{fig:partest2},~\ref{fig:partest3} represent histograms for the best individual genes (user-controlled parameters) values during their evolution in population with \ref{fig:partest1}) 20 individuals; \ref{fig:partest2}) 30 individuals; \ref{fig:partest3}) 50 individuals. All tests were performed under $p_\mathrm{cross} = 0.8$, $p_\mathrm{mut} = 0.1$ and single-point crossover. In this case we do not distinguish between different de-dispersion steps or downsampling factors. The diversity in explored parameter ranges and their quantities relates to the diversity of degenerate end solutions. Parameter histograms for larger populations are more scarce -- as GAs with larger populations need more time to initialize individuals, less time is spent on evolution of the best individual and exploration of its better genes.

\end{document}